\documentclass[submitting, floatfix]{nst}
\usepackage{subfigure,dcolumn}
\usepackage[T2A,T1]{fontenc}
\usepackage[english]{babel}
\usepackage{color}
\usepackage{listings}
\usepackage{natbib}
\let\footnote\relax

\let\textcite\relax

\let\citeauthor\relax
\let\citeyear\relax
\expandafter\let\csname
ver@natbib.sty\endcsname\relax

\let\linenumbers\relax

\usepackage[
    backend=biber,
    bibencoding=utf8,
    style=numeric-comp,
    giveninits=true,
    autolang=other, 
    sorting=none,
    doi=true, isbn=false, eprint=false, url=false, date=year, 
    maxcitenames=3,
    mincitenames=3,
    maxbibnames=3,
    minbibnames=3,
]{biblatex}

\usepackage{csquotes}
\renewbibmacro{in:}{}
\begin{document}

\renewcommand{\bibliography}[1]{}

\title{Improving the machine learning based vertex reconstruction for large liquid scintillator detectors with multiple types of PMTs}

\author{Zi-Yuan Li}
\affiliation{School of Physics, Sun Yat-Sen University, Guangzhou 510275, China}
\author{Zhen Qian}
\affiliation{School of Physics, Sun Yat-Sen University, Guangzhou 510275, China}
\author{Jie-Han He}
\affiliation{School of Physics, Sun Yat-Sen University, Guangzhou 510275, China} 
\author{Wei He}
\affiliation{School of Physics, Sun Yat-Sen University, Guangzhou 510275, China}
\author{Cheng-Xin Wu}
\affiliation{School of Physics, Sun Yat-Sen University, Guangzhou 510275, China} 
\author{Xun-Ye Cai}
\affiliation{School of Physics, Sun Yat-Sen University, Guangzhou 510275, China}
\author{Zheng-Yun You}
\email[Corresponding author, ]{youzhy5@mail.sysu.edu.cn}
\affiliation{School of Physics, Sun Yat-Sen University, Guangzhou 510275, China}
\author{Yu-Mei Zhang}
\email[Corresponding author, ]{zhangym26@mail.sysu.edu.cn}
\affiliation{Sino-French Institute of Nuclear Engineering and Technology, Sun Yat-Sen University, Zhuhai 519082, China}
\author{Wu-Ming Luo}
\email[Corresponding author, ]{luowm@ihep.ac.cn}
\affiliation{Institute of High Energy Physics, Chinese Academy of Sciences, Beijing 100049, China}

\begin{abstract}
Precise vertex reconstruction is essential for large liquid scintillator detectors.
A novel method based on machine learning has been successfully developed to reconstruct the event vertex in JUNO previously.
In this paper, the performance of machine learning based vertex reconstruction is further improved by optimizing the input images of the neural networks.
By separating the information of different types of PMTs as well as adding the information of the second hit of PMTs, the vertex resolution is improved by about 9.4 \% at 1~MeV and 9.8 \% at 11~MeV, respectively.   

\end{abstract}

\keywords{
JUNO, Liquid scintillator detector, Neutrino experiment, Vertex reconstruction, Machine Learning}

\maketitle

\section{Introduction}
\label{sec:intro}
Liquid scintillator (LS) detectors have been widely used in neutrino experiments such as KamLAND\cite{Gando:2013nba}, Borexino\cite{borexino}, Daya Bay\cite{DayaBay:2012fng}, Double Chooz\cite{doublechooz} and RENO\cite{reno}.
These experiments have made significant achievements in neutrino physics during the past few decades.
As the next generation LS detector, JUNO\cite{JUNO:2022hxd} will continue to probe the mysteries of neutrinos. 
The primary goal of JUNO is to solve the neutrino mass ordering puzzle by precisely measuring the energy spectrum of reactor neutrinos. JUNO will also be the first experiment to measure 
three of the neutrino oscillation parameters to sub-percent level. In addition, JUNO will cover a wide range of other physics topics like supernova neutrinos, solar neutrinos, atmospheric neutrinos, etc. 
In the O(1) MeV regime, particularly for reactor neutrinos, one of the main challenges for JUNO is the precise vertex and energy reconstruction of positrons, which are the prompt signals of neutrino inverse beta decay interactions.
Precise vertex reconstruction will largely help the event selection such as the fiducial volume cut and the distance cut between the correlated prompt positron and delayed neutron capture signals for reactor neutrinos. Moreover, it will also correct the energy non-uniformity, which is one of the main contributors to the energy resolution\cite{wu:2019,Huang:2021baf}.
Unlike Water Cherenkov detectors such as Super-K~\cite{PhysRevD.83.052010}(Hyper-K)~\cite{Hyper-Kamiokande:2018ofw} which can utilize Cherenkov rings, or Time Projection Chamber detectors such as DUNE~\cite{DUNE:2020lwj} which can provide track information, LS detectors have neither
clear rings nor tracks, making the vertex reconstruction relatively more challenging.

The energy deposition of positrons in LS usually consists of two parts: the kinetic energy part is roughly point-like, while the annihilation part produces two gammas and their energy is deposited within a few centimeters rather than a point. As the positron energy increases it behaves more and more like a point source. 
Previously a maximum likelihood method~\cite{Li_2021} was developed to reconstruct the vertex of positrons--the energy deposition center to be more precise using mainly the time information of the first photon hit of photo-multiplier tubes (PMTs) together with the scintillation timing profile of LS.
Ref.~\cite{Li_2021} also showed that the charge distribution of all the PMTs is sensitive to the vertex of the positron, especially near the detector boundary. A novel method~\cite{QIAN2021165527} based on machine learning was applied to JUNO reconstruction as well. 
Each PMT was treated as a pixel and the ensemble of charge or first hit time of tens of thousands of PMTs formed an image. These images were fed into neural networks to reconstruct the positron vertex. 
In Ref.~\cite{QIAN2021165527}, different neural network models such as VGG~\cite{VGG} and ResNet~\cite{Resnet} were tested and compared, the detailed structures of these models were also slightly optimized to get better reconstruction performance. 
In this paper, we will continue to explore the application of machine learning to the vertex reconstruction in large LS detectors, using JUNO as an example. 
Instead of optimizing the neural network models, we will focus on the input data, and try to optimize the input images to the networks for better vertex reconstruction performance.

The rest of this paper is structured as follows: Sec.~\ref{sec:detector} briefly describes the JUNO detector and Sec.~\ref{sec:samples} lists all the data samples used. 
Sec.~\ref{sec:optimizationI} presents one optimization of the input images by separating the different types of PMTs.
Sec.~\ref{sec:optimizationII} shows the other optimization by including the information of the second photon hit of PMTs. 
Finally, Sec.~\ref{sec:summary} gives the summary.   

\section{JUNO central detector}
\label{sec:detector}
The Central Detector (CD) of JUNO is made up of an acrylic sphere containing 20,000 tons of LS. 
The acrylic sphere is supported by a stainless-steel shell submerged in pure water. 
About 17,600 20-inch PMTs and 25,600 3-inch PMTs are installed on the stainless-steel shell to collect photons. 
The details of the JUNO CD can be found in Ref.~\cite{JUNO:2022hxd,juno}.
In this section, additional information about JUNO PMTs will be discussed. 
On one hand, JUNO is a good example of using multiple types of PMTs. 
On the other hand, all the event information such as vertex or energy are reconstructed from the PMTs signals and their precision heavily relies on the characteristics of the PMTs.  

PMTs are widely used in neutrino and other experiments for photon detection. 
As the scale of detectors increases and the requirement on the measurement precision becomes more stringent, these experiments have driven the R\&D of PMTs in return.
For small and medium scale detectors such as Daya Bay~\cite{DayaBay:2012fng}, Borexino~\cite{Agostini:2019dbs}, and SNO+~\cite{sno+}, 8-inch PMTs are utilized. 
Meanwhile, large scale detectors such as Kamiokande, Super-K, KamLAND, JUNO and Hyper-K unexceptionally use 20-inch PMTs, given their best performance-to-price ratio. 

To date there are mainly two types of 20-inch PMTs on the market for experimental usage, one is the Dynode PMT from Hamamatsu company and the other is from NNVT company (North Night Vision Technology Co. Ltd.) with a novel Micro-channel Plate (MCP) design. 
Each type has its own specifications. 
The physics potential highly depends on the performance of PMTs. 
However, it is non-trivial to choose the most appropriate PMTs for an experiment after taking into account not only the PMT characteristics, but also the cost and risk. 
Ref.~\cite{WEN2019162766} presented an interesting and quantitative strategy of PMT selection for large detectors. 
In the case of JUNO, 12,612 MCP PMTs and 5,000 Dynode PMTs will be installed. 
Tab.~\ref{tab:PMT_comp} shows the comparison between these two types of PMTs for the parameters which are relevant to vertex reconstruction.
\begin{table}[!h]
\caption{Comparison between the two types of PMTs in JUNO. Only parameters relevant to vertex reconstruction are listed.}
\begin{center}
\begin{tabular}{ccc}
\toprule
 & \textbf{Dynode} & \textbf{MCP} \\
 \midrule
Detection efficiency [$\%$] & 28.4  & 30.1  \\
Dark noise rate [kHz] & 15.3  & 29.6 \\ 
Charge resolution [$\%$]& 27.9  & 32.9 \\
Transit time spread [ns]  & 2.8  & 12.0 \\
\bottomrule

\end{tabular}
\end{center}
\label{tab:PMT_comp}
\end{table}

The MCP PMTs have slightly better photon Detection Efficiency with an average value of 30.1\%, with respect to 28.4\% for Dynode PMTs. 
The intrinsic charge resolution is slightly better for Dynode PMTs. 
The average dark noise rate for MCP PMTs is about twice of that for Dynode PMTs. 
Another key difference is the transit time spread (TTS), which is 2.8~ns for Dynode PMTs and 12~ns for MCP PMTs, respectively. 
As a result, Dynode PMTs have much better time resolution compared to MCP PMTs. 
JUNO deliberately chose to use about 28.3\% Dynode PMTs in order to achieve better vertex resolution. 
In addition to the 20-inch (large) PMTs, JUNO will also install 25,600 3-inch (small) PMTs as mentioned previously.
In principle, the small PMTs could be used to improve the reconstruction performance. 
However, due to their small geometrical coverage ($\sim$ 3\%) for photons, they are not considered in this paper.

\section{Monte Carlo Samples and Reconstruction Method}
\label{sec:samples}
To study the vertex reconstruction of positron events in JUNO with machine learning techniques, different positron samples are prepared and the relevant information is summarized in Tab.~\ref{tab:PhySampleInfo}. 
The training sample is used to train the machine learning models.
The training process usually requires a huge amount of events. 
Given the large volume of JUNO, 5 million Monte Carlo (MC) events are simulated as the training sample.
The vertices of these events are uniformly distributed in the whole detector volume and their kinetic energy range from 0~MeV to 10~MeV.
Eleven sets of testing samples with kinetic energy $E_k$~= (0, 1, 2, ..., 10)~MeV are used to evaluate the performance of the vertex reconstruction. 
These testing samples are uniformly distributed in the whole detector volume as well. 
The statistics for each testing sample is 0.5 million.

For all these samples, the detector simulation is performed with the JUNO offline software based on Geant4~\cite{Geant4}, including LS properties and optical processes of photon propagation~\cite{ABUSLEME2021164823,lsopmdl}. The event display software~\cite{You:2017zfr,Zhu_2019} dedicated to JUNO can be used to dynamically display the entire process.
Realistic detector geometry such as the arrangement of the PMTs and the supporting structures is also deployed\cite{zhang2020method,Li:2018fny}. 
Unlike Ref.\cite{QIAN2021165527} which does not include the charge smearing and waveform of PMTs, the MC data samples in this paper have gone through the full chain of detector simulation, electronics simulation, PMT waveform reconstruction and PMT calibration, making them as close to real data as possible.
Two sets of data samples referred to as the ideal and real samples are produced, in which the electronics effects such as TTS and dark noise of PMTs are disabled or enabled respectively.

\begin{table}[!h]
\centering
\caption{List of the positron samples used for CNN training and testing.}
    \begin{tabular}{lccc}
\toprule
            & \textbf{Kinetic energy} & \textbf{Statistics} & \textbf{Position}\\
\midrule
     Training & uniform in [0, 10]~MeV& 5M & uniform in CD \\
     Testing & (0,1,2, ..., 10)~MeV& 500k $\times$ 11 & uniform in CD \\
\bottomrule
\end{tabular}
\label{tab:PhySampleInfo}
\end{table}

\begin{figure*}[htbp]
	\centering
	\includegraphics[width=0.9\textwidth]{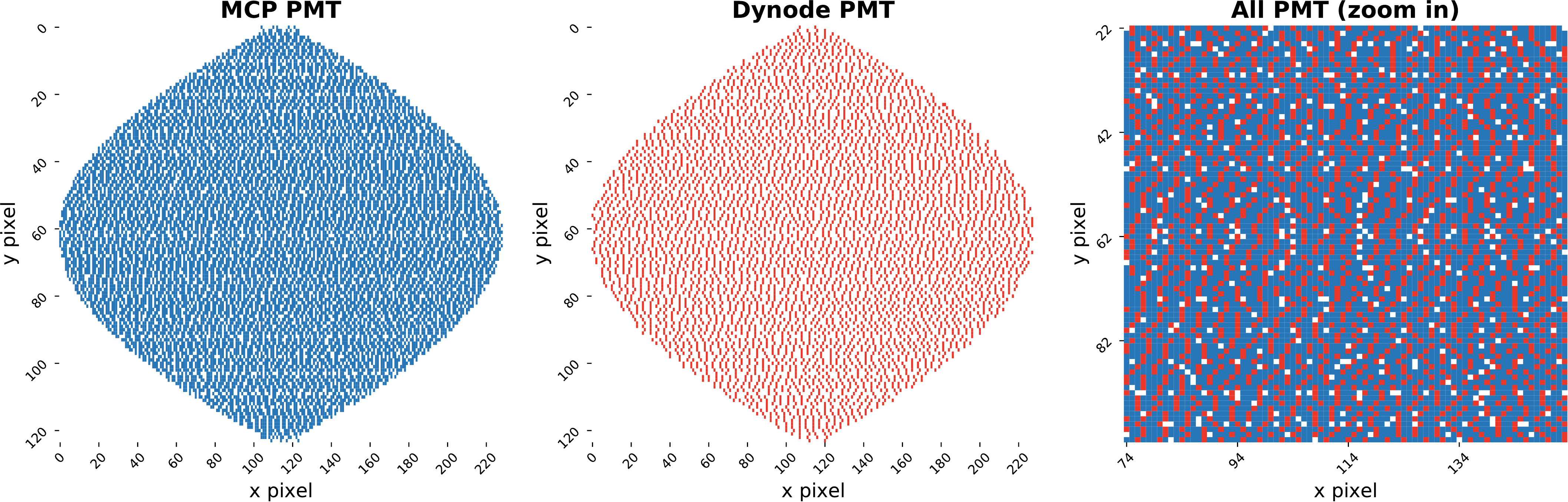}{\centering}
	\caption{2D plane projection of the PMTs. 
	The PMTs are projected to the plane image (229 $\times$ 124) based on their positions and details can be found in the context.
	The left and middle plots correspond to Dynode and MCP PMTs, respectively. The right plot shows the two types of PMTs overlaid in a small region. The white spots are empty pixels. The size of the image has been optimized to avoid any overlap of PMTs and minimize the number of empty pixels.}
	\label{fig:PMT2D}
\end{figure*}

\begin{figure*}[!ht]
	\centering
	\includegraphics[width=1.0\textwidth]{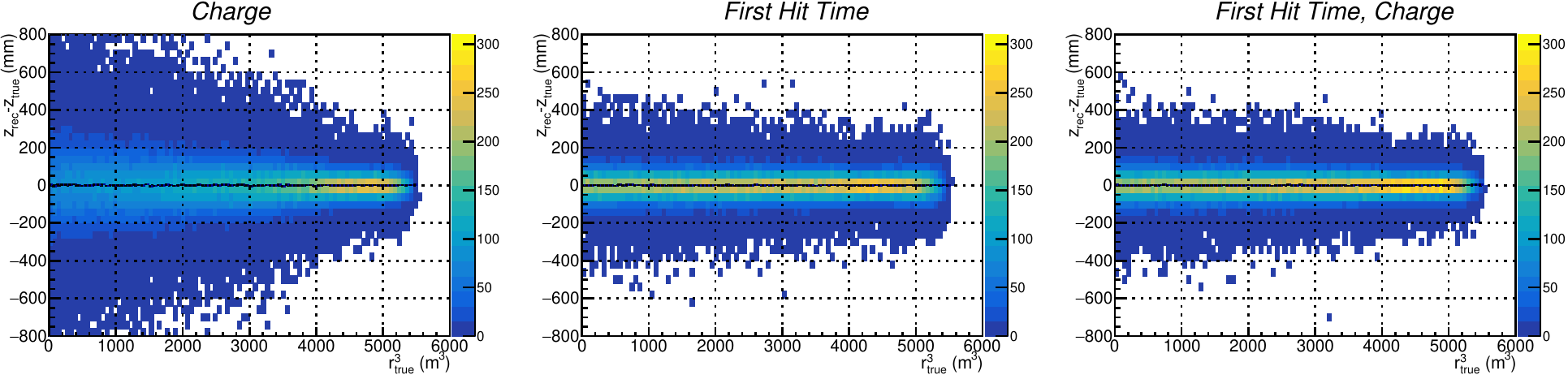}{\centering}	
	\caption{The distribution of $\delta Z$ as a function of the cubic of radius $r^3$. The left plot corresponds to Case A in which both Charge and FHT information are used.  The middle plot corresponds to Case B with FHT information only. The right plot corresponds to Case C with Charge information only. The black curve in each plot shows the vertex bias, which is close to 0 in all cases.}
	\label{fig:charge_time_bias}
\end{figure*}

The vertex reconstruction method in this paper is inherited from Ref.\cite{QIAN2021165527}. 
All the PMTs on the spherical stainless-steel shell are projected to a 2D plane based on their positions as shown in Fig.~\ref{fig:PMT2D}. The PMTs will be installed ring by ring from the bottom of the CD to the top. For each PMT, its Y pixel number corresponds to its ring number, its X pixel number is calculated with:
\begin{eqnarray}
    \begin{aligned}
         &X_{\text{pixel}} = \left[N_{\text{eff}} \cdot \frac{\arctan(x/y)}{\pi} \right]+ \frac{ N_{\text{max}}}{2}, \\
        &N_{\text{eff}} =  \left[N_{\text{max}} \cdot  \frac{\sqrt{R^2-z^2}}{R}\right],
    \end{aligned}
    \label{eq:xyz}
\end{eqnarray}
where $x,y,z$ is the global position of PMT, $R$ is the radius of the central detector and $N_{\text{max}} = 229$ has been optimized to avoid overlap of PMTs and minimize the number of empty pixels.
The charge or time information of all the PMTs for any event will form an image whose pattern varies for different event vertex. 
These images (or channels in CNN jargon) are then fed into a convolutional neural network (CNN) as inputs and the output will be the event vertex. 
After a specific CNN model is trained, it can be used to reconstruct the event vertex.
In Ref.\cite{QIAN2021165527} various CNN models were compared, VGG and ResNet were found to give the best performance.
The structures of the neural networks in these two models are also slightly optimized and tailored to the specific requirements of JUNO. 
Thus the "J" in VGG-J and ResNet-J stands for JUNO. 
These two models roughly have the same performance for vertex reconstruction, VGG-J is chosen for all the studies in this paper simply due to its faster training process.

\section{Optimization of Input Images by Separating Different Types of PMTs}
\label{sec:optimizationI}

\subsection{Charge vs Time}
One follow-up question in JUNO reconstruction from Ref.~\cite{QIAN2021165527} is the relative importance of charge and time information of PMTs in the vertex reconstruction. 
The following three cases were tested with the same data samples as well as the same CNN model. 
In Case A both the charge and First Hit Time (FHT) images are used, while in Cases B and C only the FHT image or the charge image is used, respectively.
One thing to note is that since the input to the CNN is different, the model is retrained in each case.

\begin{itemize}
	\item Case A: both Charge and FHT images are used
	\item Case B: only FHT image is used
	\item Case C: only Charge image is used
\end{itemize}

Given the rough spherical symmetry of the JUNO CD, results of the vertex reconstruction are quite similar for the X, Y and Z components of the vertex, as shown by Figure 11 from Ref.~\cite{QIAN2021165527}. So 
only the Z component will be presented throughout this paper.
We denote $\delta Z$ as the difference between the reconstructed $Z_{rec}$ and the true $Z_{edep}$  (from the energy deposition center). 
After fitting the distribution of  $\delta Z$ with a Gaussian function, the Gaussian mean and standard deviation are defined as the vertex bias and vertex resolution, respectively.  

\begin{figure}[htbp]
	\centering
	\includegraphics[width=0.42\textwidth]{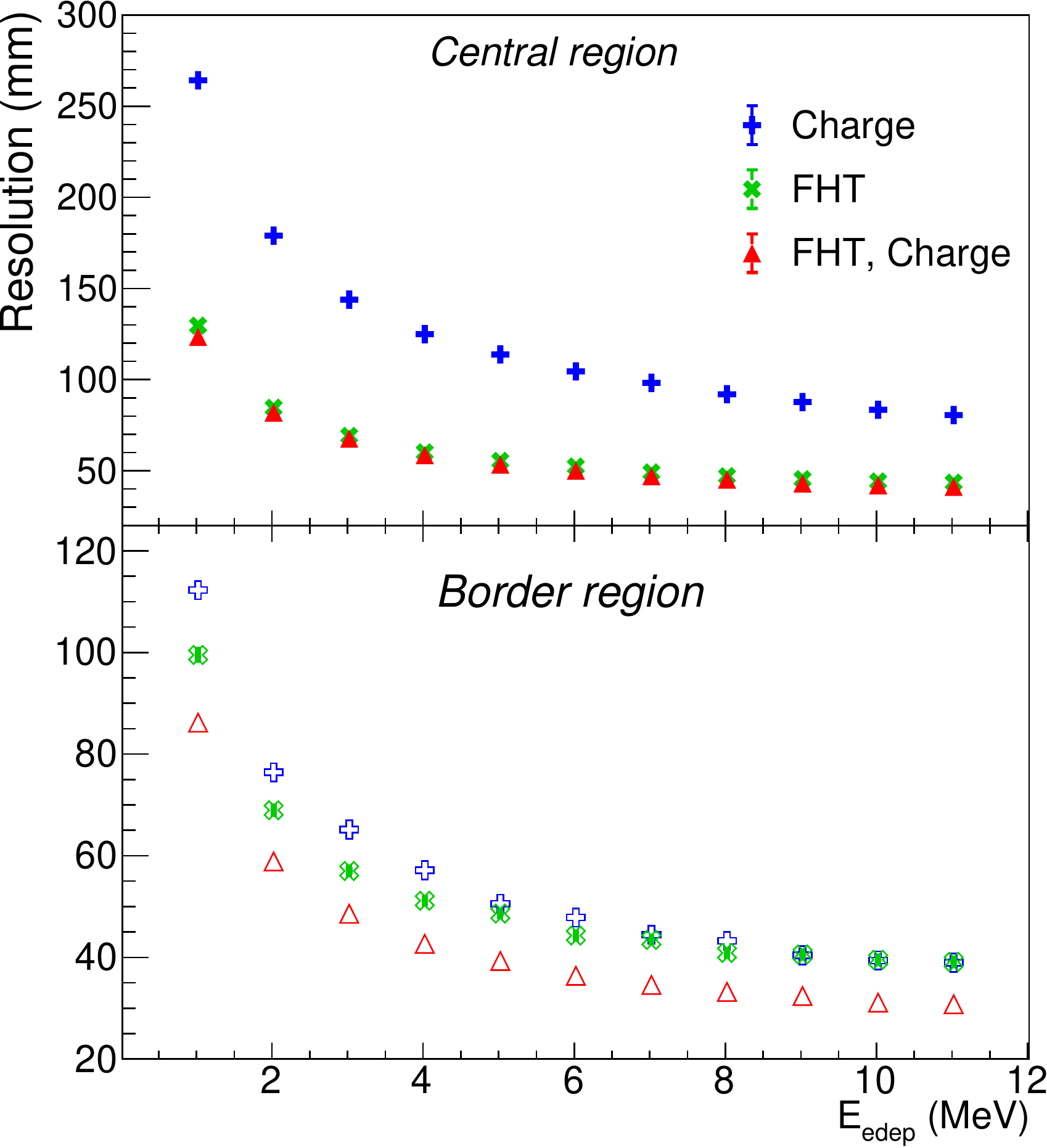}{\centering}
    \caption{Energy dependence of the vertex resolution for the three cases (Charge, FHT, FHT \& Charge). The top and bottom panels correspond to the central ($r^3$ $<$ 4000 $\text{m}^3$) and border regions ($r^3$ $>$ 4000 $\text{m}^3$) of the detector, respectively.}
	\label{fig:charge_time_res}
\end{figure}

Fig.~\ref{fig:charge_time_bias} shows how $\delta Z$ changes with respect to the cubic of radius $r^3$ for positrons from the testing samples in the three cases. 
In all cases the vertex bias represented by the black curve in each plot is close to zero in the whole detector. 
This statement also holds for all the later cases in this paper. 
Thus we will not show the vertex bias anymore hereafter. 
Meanwhile, the vertex resolution is better in the border region with $r^3 >$ 4000 $\text{m}^3$ indicated by the narrower spread of $\delta Z$. 
Overall FHT information is much more powerful to constrain the vertex in the central region with $r^3 <$ 4000 $\text{m}^3$ compared to the charge information, but in the border region their performance of the vertex reconstruction are rather close. 
This can be seen more clearly from Fig.~\ref{fig:charge_time_res}, which compares the dependence of the vertex resolution on energy for the three cases in the central region and the border region, respectively. 
Fig.~\ref{fig:charge_time_res} also shows that using both charge and FHT always gives better vertex resolution compared to using charge or FHT only in both regions. 
The charge and FHT of PMTs provide complementary information and should both be used to achieve the best vertex reconstruction.

\subsection{Dynode vs MCP}
Similar to the charge vs time comparison, another question is how the two types of PMTs contribute to the vertex reconstruction. 
To address this question, the vertex reconstruction was performed using different types of PMTs as listed below.
Again the same data samples and CNN model were used here, the inputs to the CNN include both the FHT and charge images and the CNN is retrained for each case. 
\begin{itemize}
	\item Case 1: only MCP PMTs are used
	\item Case 2: only Dynode PMTs are used
	\item Case 3: both types of PMTs are used
\end{itemize}

\begin{figure}[htbp]
	\centering
	\includegraphics[width=0.42\textwidth]{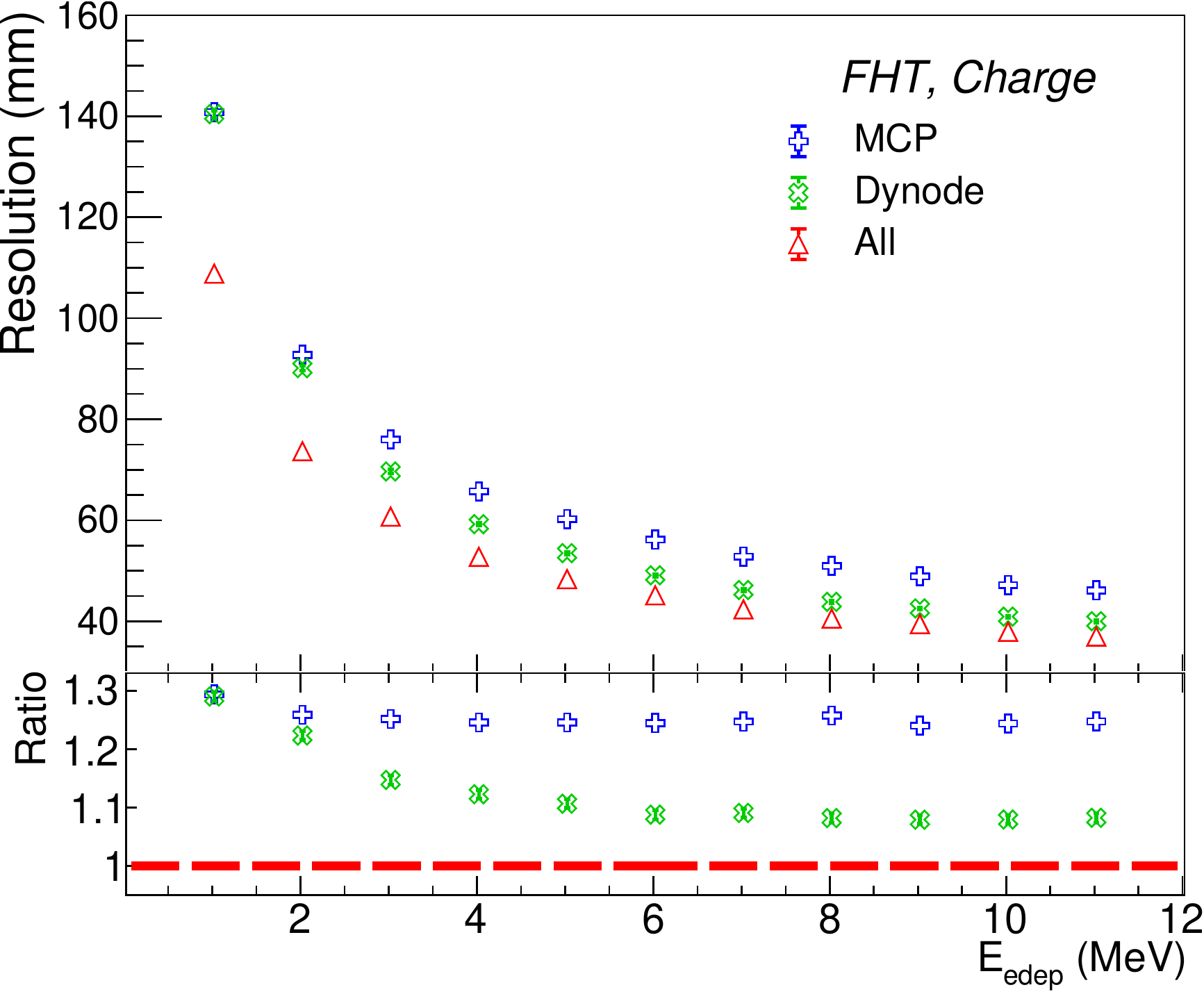}{\centering}	
	\caption{Comparison of the vertex resolution among the three cases in which different types of PMTs are used.
Blue represents Case~1 where only MCP PMTs are used. Green represents Case~2 where only Dynode PMTs are used. 
Red represents Case~3 where both types of PMTs are used. The bottom panel shows the ratio of Case~X/Case~3.}
	\label{fig:pmt_comp_res}
\end{figure}

The comparison of the vertex resolution among the three cases is shown in Fig.~\ref{fig:pmt_comp_res}.
Blue dots correspond to Case~1 where only MCP PMTs are used. Green dots correspond to Case~2 where only Dynode PMTs are used. 
Red dots represent Case~3 where both types of PMTs are used. 
Although the total number of Dynode PMTs is less than half of MCP PMTs, Dynode PMTs have much better time resolution because of much smaller TTS. 
As a result, in comparison to Case~1 with only MCP PMTs, Case~2 with only Dynode PMTs gives better vertex resolution nearly across the whole energy range except the lowest energy point. 
This is more or less consistent with what we have learned from traditional vertex reconstruction algorithms~\cite{Li_2021}, the vertex resolution is approximately proportional to $\sigma_{TTS}$/$\sqrt{N}$, where N is the number of fired PMTs and $\sigma_{TTS}$ is the time resolution of PMTs.

Naively one would have thought Case~3 could have much better vertex reconstruction performance comparing to both Case~1 and Case~2, since the information of both types of PMTs are used. 
However this is not true. 
In the high energy region, Case~3 actually only has slightly better vertex resolution than Case~2. Although in principle more information should provide additional constraint to help improve the vertex reconstruction, how the information is utilized also matters a great deal. 
An analogy could be made for Case~3 from above. 
Imagine that we have two cameras with drastically different resolution. 
If we use them to take an image of the same object and then simply overlay the two images on top of each other to forge a combined image. 
The camera with much worse resolution might not help to improve the quality of the combined image.
On the contrary, it might make the combined image more fuzzy and possibly even degrade its quality.  MCP PMTs have much worse time resolution compared to Dynode PMTs, overlaying their FHT images in the same one could confuse the network and result in marginal improvement on the reconstruction performance. 
As for the low energy region, the number of fired PMTs becomes more important. 
Using both types of PMTs leads to much better vertex resolution.

\subsection{Separation of input images of PMTs}
\label{sec:channel}
In order to achieve the best performance of the vertex reconstruction, the information of both types of PMTs should be used. 
However, mixing the information of different types of PMTs together might not be the optimal way as one can see from the above section. 
In the traditional method of the vertex reconstruction from Ref.~\cite{Li_2021}, the two types of PMTs 
are handled separately with different residual time PDFs due to different TTS. 
Following the same strategy, the FHT information should be separated into two images (or channels), one for each type of PMTs.
Given that the charge resolution is also different for Dynode and MCP PMTs, the charge information
could be segregated as well. 
To test the performance of this new strategy, a few scenarios were considered as listed below:
\begin{itemize}
\item Default case: both the charge and FHT information are mixed together for the two types of PMTs and the input to VGG-J includes one charge image plus one FHT image.
\item Partially separated case: the FHT information is segregated by PMT types and the there are three input images.
\item Fully separated case: both the charge and FHT images are separated, resulting in four input images. 
\end{itemize}

\begin{figure}[htbp]
	\centering
	\includegraphics[width=0.42\textwidth]{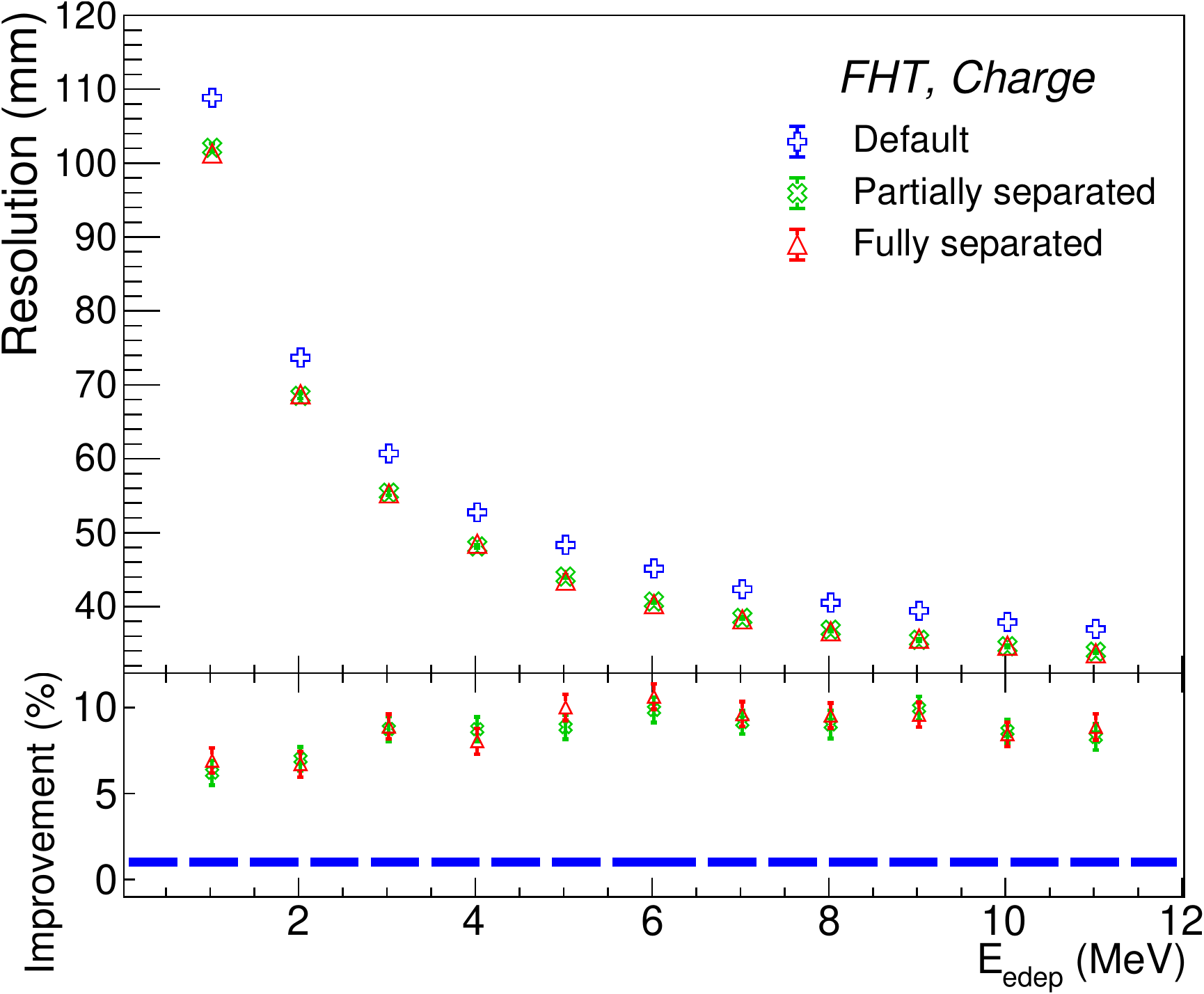}{\centering}	
	\caption{Comparison of the vertex resolution among the three cases with different input images to VGG-J. Blue represents the default case where mixed charge and FHT images are used. Green represents the partially separated case where the FHT image is separated. Red represents the fully separated Case where both FHT and charge images are segregated. The bottom panel shows the improvement with respect to the default case.}
	\label{fig:channel_comp}
\end{figure}

The VGG-J model is retrained for each case and the performance of the vertex reconstruction is evaluated and compared. Fig.~\ref{fig:channel_comp} shows the comparison of the vertex resolution for the three cases. 
The red, blue and green dots represent the default, partially separated and fully separated cases respectively. 
After separating the FHT information by PMT types, a large improvement is observed across the entire energy range with respect to the default case. 
For example, at 1~MeV the vertex resolution decreases from 111~mm to 102~mm and at 11~MeV it decreases from 37~mm to 34~mm. 
Further separation of the charge information also leads to better performance with respect to the partially separated case, however, the improvement is small. 
For example, the vertex resolution at 1~MeV only improves from 102~mm to 101~mm. 
This is what one would expect given the large TTS difference of the two types of PMTs. 
While on the other hand, their charge resolution is not so different. 

For the JUNO CD, 
although the small PMTs are not included in this paper for simplicity, it is straightforward to add their information for the vertex reconstruction using the same strategy from above.
When there are multiple types of PMTs in a detector, the best strategy would be to utilize their information separately, especially when the characteristics of different types of PMTs are very different. 
This is certainly true for vertex reconstruction as demonstrated in this paper, it might also be applicable to other tasks in general. 
Reusing the camera analogy, each type of PMTs actually forms an independent camera or sub-detector, their images or measurements should be taken separately and then combined to achieve the optimal performance.

\begin{figure*}[!ht]
	\centering
	\includegraphics[width=1\textwidth]{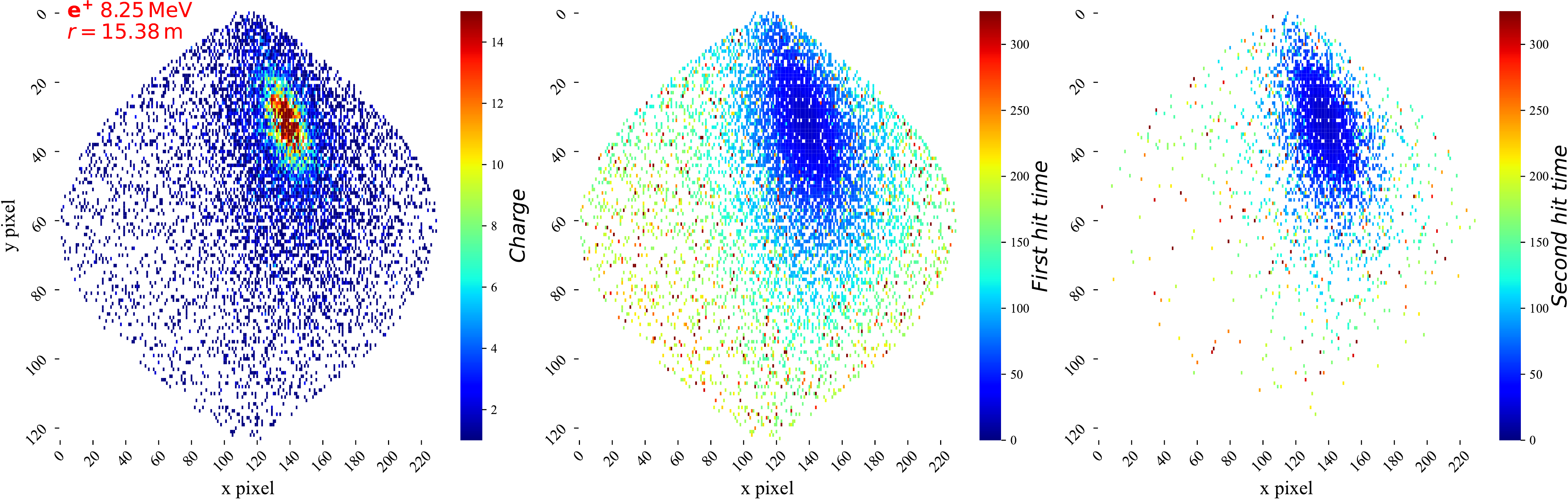}{\centering}	
	\caption{Images of PMT charge and time information for a positron with E = 8.25~MeV and r = 15.38~m in the JUNO CD. The left, middle and right plots show the images of charge, FHT and SHT, respectively. The time window for photon counting is 1250~ns for PMTs.}
	\label{fig:eventDisplay}
\end{figure*}
\section{Addition of Second Hit}
\label{sec:optimizationII}
In Ref.~\cite{QIAN2021165527} as well as all the studies above, the inputs to the CNN models are only limited to the total charge and the first hit time of each PMT. 
It is possible that more than one photon will hit a PMT and the possibility is both energy and vertex dependent. 
As the event energy increases, more photons will be emitted, consequently all PMTs are more likely to detect more photons. 
On the other hand, when the event vertex gets closer to the border of the detector, those PMTs near the event vertex will probably receive more photons.  

\begin{figure}[htbp]
	\centering
	\includegraphics[width=0.42\textwidth]{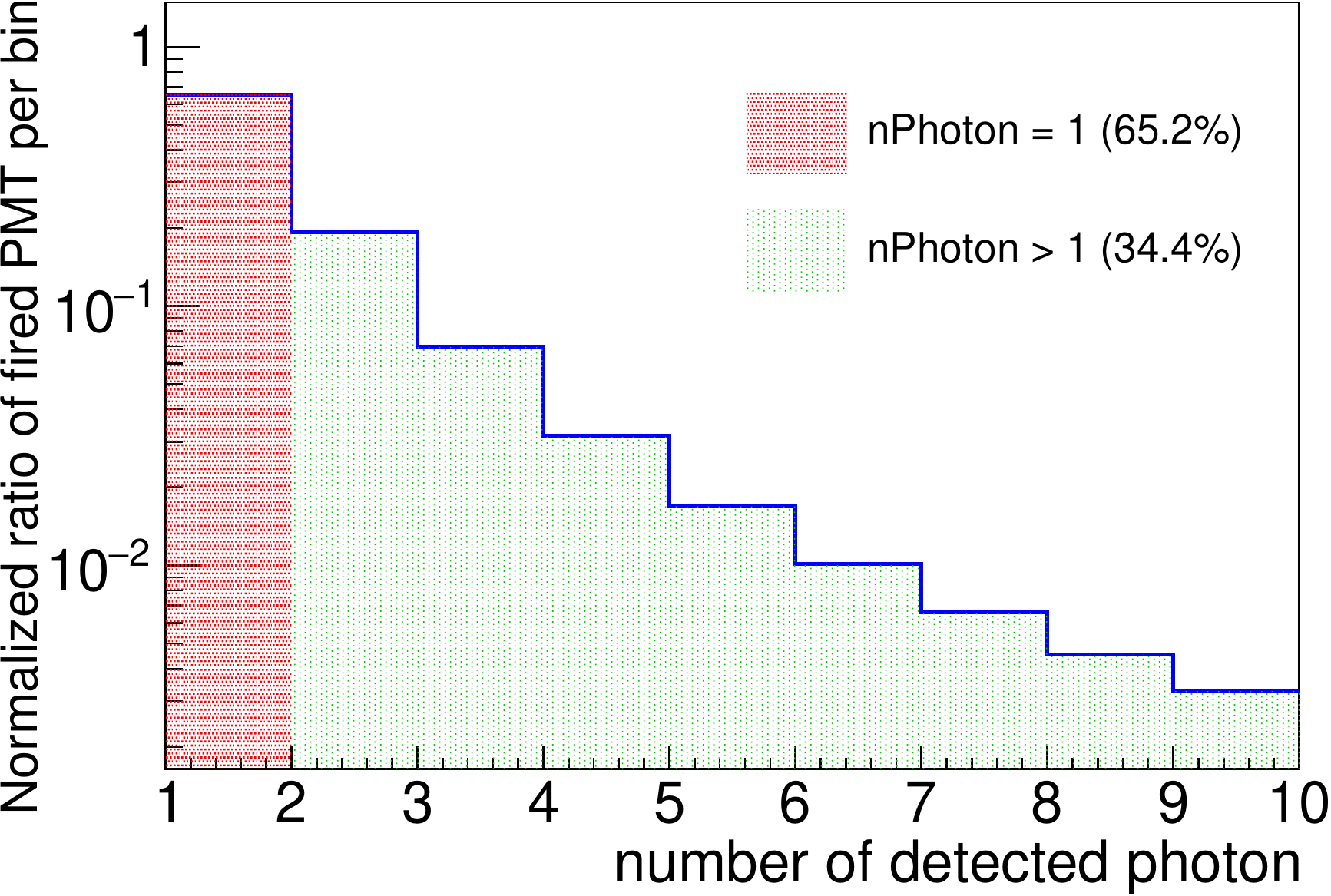}{\centering}	
	\caption{Normalized distribution of the true number of detected photons for all PMTs in detector simulation for the training sample.}
	\label{fig:nPE_distribution}
\end{figure}

Given the large number of PMTs (about 17,600 in total) for the JUNO CD and the rough light yield of 1,300 photons per MeV, for positrons with energy less than 11~MeV, most of the PMTs will receive zero or one photon. 
About one third of those fired PMTs will detect more than one photon on average for all the events in the training datasets as shown in  Fig.~\ref{fig:nPE_distribution}. 
The fraction of fired PMTs which detect three or more photons will drop sharply since the number of detected photons for each PMT obeys Poisson distribution with small mean values. 

In principle, all these later photon hits of PMTs also contain information about the event vertex.
However, we will only consider the second hit time (SHT) of PMTs in this paper. 
If the addition of the SHT does not improve the vertex reconstruction, it is unnecessary to include the third or later hits since their fraction is even smaller.

\subsection{Ideal case without TTS and dark noise}
\begin{figure}[htbp]
	\centering
	\includegraphics[width=0.42\textwidth]{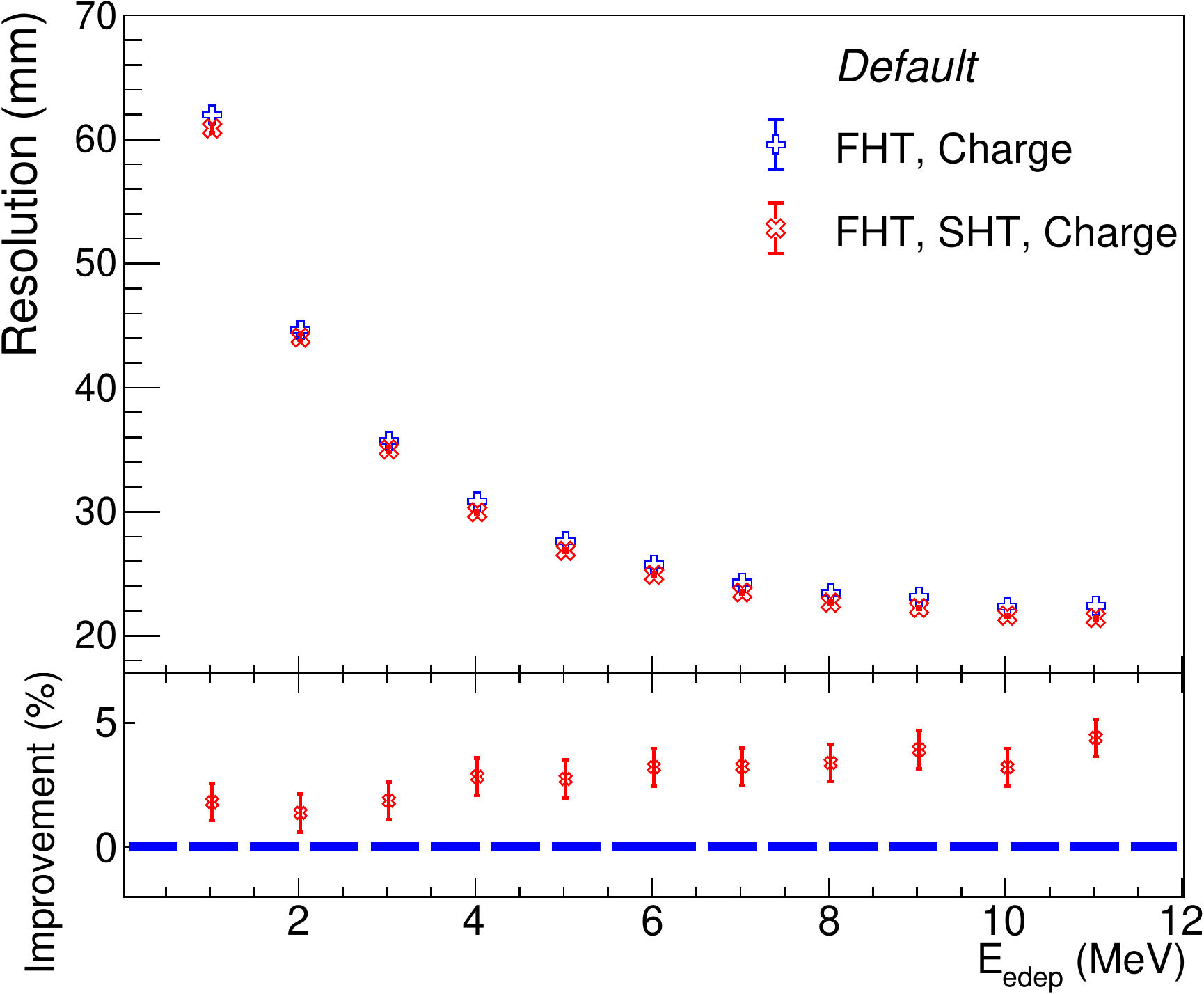}{\centering}	
	\caption{Comparison of the vertex resolution with and without using the SHT information in the ideal case, represented by the red and blue dots respectively. The bottom panel shows the improvement by adding SHT, which is more pronounced as the energy increases.}
	\label{fig:SHTideal}
\end{figure}

We start from the ideal case in which the electronics effects such as PMT TTS and dark noise are turned off.
In this scenario, the two types of PMTs could basically be treated as the same.
Fig.~\ref{fig:eventDisplay} shows the PMT images for a positron with E = 8.25~MeV and r = 15.38~m in the JUNO CD in the ideal case. 
The left, middle and right plots correspond to the images of charge, FHT and SHT, respectively. 
The projection of PMTs to the 2D plane is the same as that in Fig.~\ref{fig:PMT2D}.
By comparing the FHT and SHT images, similar patterns are obvious, those PMTs closer to the vertex in the FHT images are more likely to detect two or more photons and contribute to the SHT image.
Presumably the SHT information could add additional constraints on event vertex. 
This can be easily checked by adding the SHT image to the VGG-J model. 
The reconstruction results are plotted in Fig.~\ref{fig:SHTideal}, which are represented by the red dots.
The reconstruction results without using the SHT image are also drawn as the blue dots for comparison. In general, adding SHT improves the vertex resolution and it is more pronounced as the energy increases. 
At 1~MeV the vertex resolution improves by about 1.6\% from 62~mm to 61~mm after adding the SHT image, while at 11~MeV it improves by about 4.3\% from 23~mm to 22~mm.
This is consistent with our expectation since the fraction of PMTs with SHT becomes larger as the energy increases.

\subsection{Realistic case with TTS and dark noise}
In the above section, we demonstrate that the SHT information is also useful for vertex reconstruction in the ideal scenario.
In reality TTS and dark noise of PMTs must be taken into account. 
In Ref.~\cite{Li_2021}, their impact on vertex reconstruction have been studied exclusively. 
The dominant effect comes from TTS since it largely degrades the resolution of FHT. 
On the other hand, the contribution of PMT dark noise to FHT is small if its rate is not too high. Thus its impact on vertex reconstruction is relatively small in Ref.~\cite{Li_2021} where only FHT is used. 
Since PMT dark noise occurs randomly in time, it does not contain any information about the event vertex. 
These additional PMT hits from dark noise will contaminate the real photon hits and should be removed. 
However, it is not easy to discriminate all the dark noise hits from the real photon hits, and this task is out the scope of this paper. 
For photons originating from the same particle and arriving at the same PMT, their corresponding 
PMT hits have strong temporal correlation. 
This correlation could be used to partially remove dark noise contribution, particularly for later hits.
We require the difference between FHT and SHT to be less than 300~ns. 
This simple cut on the SHT has been optimized so that 98.9\% real photon hits are kept while 48\% dark noise hits are rejected for SHT. 
After applying this cut, the average fraction of fired PMTs with SHT for all the events in the training dataset will decrease from about 36.3\% to 33.8\%, which is close to 33.3\% in the ideal case without dark noise.
Meanwhile, for the PMTs with SHT, the fraction of PMTs containing dark noise hits decreases from about 11.3\% to 4.4\%, similar to the number of 4.9\% for PMTs with FHT.

\begin{figure}[htbp]
	\centering
	\includegraphics[width=0.42\textwidth]{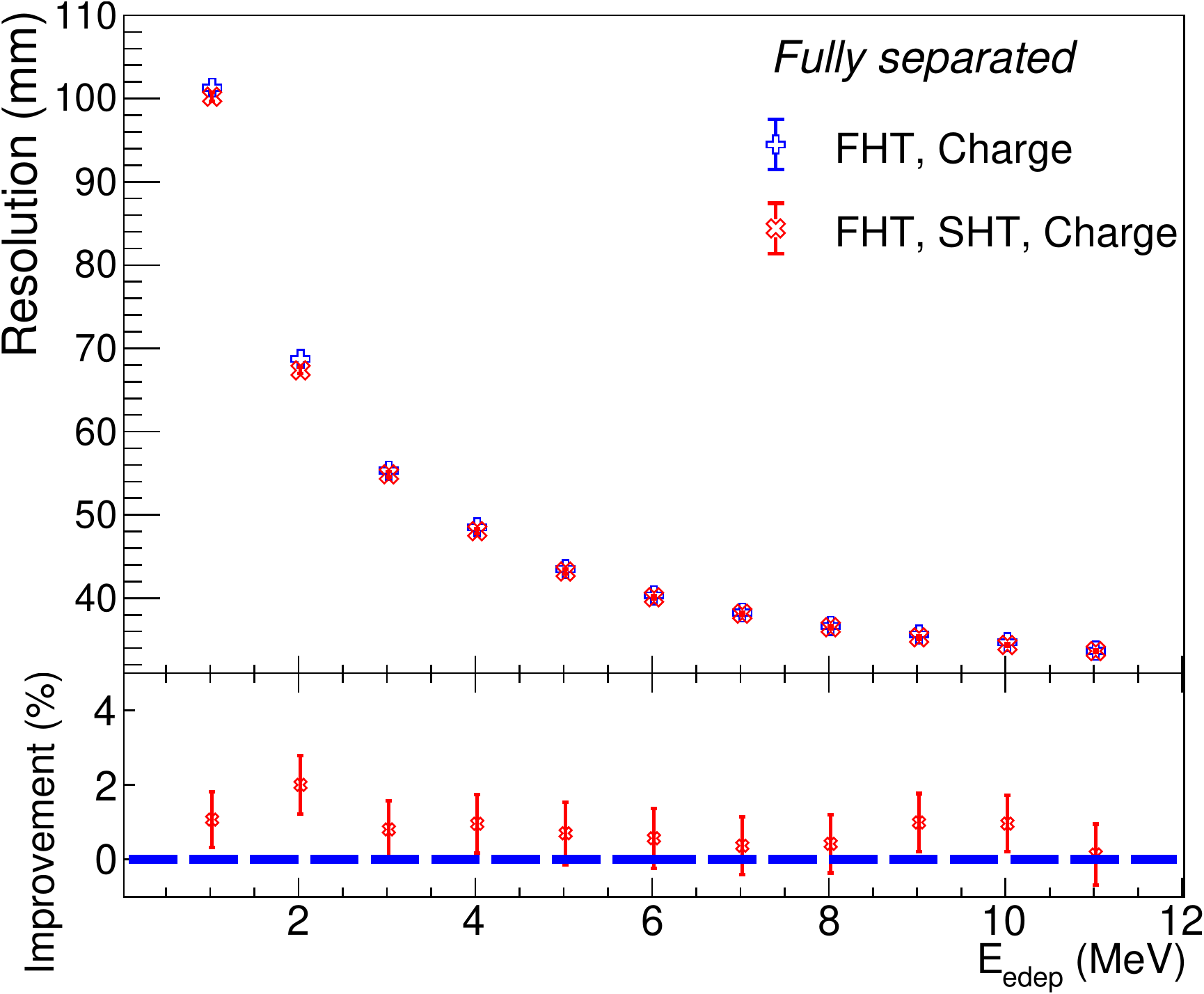}{\centering}	
    	\caption{Comparison of the vertex resolution with and without using the SHT information in the realistic case, represented by the red and blue dots respectively. The bottom panel shows the improvement by adding SHT, which is about 1\% in this case.}
	\label{fig:SHTreal}
\end{figure}

Given that TTS is no longer zero in the realistic case, the information of two types of PMTs needs to be separated in order to achieve the best performance as shown in Sec.~\ref{sec:channel}. 
On top of the charge and FHT images of both types of PMTs, the two additional SHT images are also fed into VGG-J, accounting for 6 images in total. 
Fig.~\ref{fig:SHTreal} compares the vertex resolution with and without using the SHT information in the realistic case. 
The red dots represent the fully separated case in Sec.~\ref{sec:channel}, where 4 images are used, while the blue dots represent the case where 6 images are used. 
Similar to the ideal case, adding SHT does improve the vertex resolution in the realistic case. However, the improvement is not as prominent as the ideal case, which is mainly due to the degradation of the PMT time resolution caused by TTS.

From the studies on both the ideal and realistic cases, it is clear that SHT could improve the performance of the vertex reconstruction. And the time resolution of PMTs is an essential factor. 
For any future similar detectors, we should try to reduce the TTS of PMTs. 
We also checked that the improvement by adding SHT is similar in the central and border regions for both cases.
Another thing to note is that both the charge and time information are reconstructed from PMT waveforms. 
This process will introduce additional uncertainties on both charge and time.
Thus we also need to develop better waveform reconstruction method to mitigate its impact on both the charge and time resolution of PMTs. 
Later hits might also be useful, provided that they could be well identified and reconstructed, which
tends to be difficult especially when they overlap with each other.
 
\section{Performance Summary and Discussion}
\label{sec:summary}
Two optimizations for the machine learning based vertex reconstruction have been studied in this paper, 
namely separation of the PMT information by PMT type in Sec.\ref{sec:optimizationI} and addition of the SHT information in Sec.\ref{sec:optimizationII}. 
The improvement on the vertex resolution is shown in Fig.\ref{fig:channel_comp} and Fig.\ref{fig:SHTreal}.

\begin{table}[ht]
\caption{Comparison of the vertex resolution with different options of input images. For the default option the charge and FHT information of the two types of PMTs are mixed and there are only 2 input images. For option I the information of the PMTs are separated by the PMT type, resulting in 4 input images. In option II the SHT information is also added and there are 6 input images in total. Only the vertex resolution at 1, 5 and 11~MeV are listed for the comparison. The relative improvement with respect to the previous option is also quoted in the brackets for options I and II.}
\begin{center}
\begin{tabular}{lrrr}
\hline
\hline
Option    & Default & I: PMT Separation & II: Addition of SHT\\
Images    & charge, FHT  & (charge, FHT) $\times$ 2  & (charge, FHT, SHT) $\times$ 2 \\
\hline
    1~MeV  & 110.59~mm  & 101.31~mm~(+8.4\%)  & 100.22~mm~(+1.1\%)\\
    5~MeV  &  48.70~mm &   43.50~mm~(+10.7\%)& 43.20~mm~(+0.7\%)\\
    11~MeV &  37.29~mm &  33.68~mm~(+9.7\%) & 33.64~mm~(+0.1\%)\\
\hline
\end{tabular}
\end{center}
\label{tab:Summary}
\end{table}

For the reader's convenience, Tab.~\ref{tab:Summary} summarizes the results at 1, 5 and 11~MeV.
The default case without either optimization is also shown for comparison. 
Separation of the PMT information by PMT type leads to a 10\% improvement with respect to the default case. 
Further addition of the SHT information gives another 1\% improvement.

For the machine learning based vertex reconstruction, there are still a few aspects that need to be further investigated. 
Firstly, it is a continuous process for particles to deposit energy in LS and emit photons, which is more like a video rather than an image. 
How to use this temporal information might appose new challenges. 
Meanwhile, the JUNO CD is a spherical detector, any projection of PMTs on the surface of a sphere to a 2D plane usually causes deformation and loss of symmetry and continuity. 
We could borrow the tools from astrophysics to deal with spherical images. 
Last but not the least, the robustness of machine learning techniques has to be verified, especially when there are discrepancies between the training and evaluating datasets, or the Monte Carlo simulation and real data. 
By addressing these topics in the future, we hope to achieve the best vertex reconstruction for large LS detectors with multiple types of PMTs, and consequently enhance the detector performance to increase the physics potential of new discoveries.

\section{Conclusion}
\label{sec:conclusion}
High precision vertex resolution is essential for large liquid scintillator detectors such as JUNO. 
There are quit a few publications~\cite{allhittimerec,borexino,Li_2021} on vertex reconstruction using traditional methods for  liquid scintillator detectors.
While the novel idea of the vertex reconstruction with machine learning techniques has only been applied to JUNO for the first time recently~\cite{QIAN2021165527}. 
In this paper we continue to improve the performance of machine learning based vertex reconstruction and focus on the optimization of the input images to the CNN model. 
Due to different characteristics of various types of PMTs, their information are separated rather than mixed together. 
Moreover, in addition to the FHT information of PMTs, the SHT information is also used. 
The separation of two types of PMTs leads to a noticeable improvement on the vertex resolution, about 10\% on average across the energy range of [1, 10]~MeV. 
Further addition of SHT results in roughly another 1\% improvement on average. 
These two optimizations seem to be rather simple, but they could be used as general guidelines for other detectors with multiple types of PMTs.

\section*{Acknowledgements}
This work is supported by National Natural Science Foundation of China (Grant No.11975021, 12175257, 12175321, 11675275, U1932101),
the Guangdong Basic and Applied Basic Research Foundation (2021A1515012039), 
the Strategic Priority Research Program of the Chinese Academy of Sciences under Grant No. XDA10010900, 
the National College Students Science and Technology Innovation Project,
the Undergraduate Base Scientific Research Project of Sun Yat-sen University,
and by the CAS Center for Excellence in Particle Physics (CCEPP).
We would like to thank the Computing Center of IHEP for providing the GPU clusters.

\printbibheading[heading=bibintoc]
\printbibliography[heading=none]

\end{document}